\documentclass[preprint,review,12pt]{elsarticle}

\usepackage{graphicx}
\usepackage{amssymb}
\usepackage{lineno}
\usepackage[super]{nth}

\journal{be submitted}

\begin{document}

\begin{frontmatter}


\title{A novel method to identify IMF $B_y$ influence over low-latitude ionosphere  from the correlation of magnetic field variations at nearly antipodal stations}





\author{Sumanjit Chakraborty\corref{c-d54cc1eb1ca4}}
\ead{sumanjit11@gmail.com}\cortext[c-d54cc1eb1ca4]{Corresponding author.}
\author{D. Chakrabarty}
\address{Space and Atmospheric Sciences Division, Physical Research Laboratory, Ahmedabad 380009, Gujarat, India}
\ead{dipu@prl.res.in}

\begin{abstract}

This work is directed to explore the efficacy of strong and consistent correlations between the $\Delta$X component of magnetic field at two equatorial/low-latitude stations at nearly antipodal locations during geomagnetic storms. The antipodal stations considered are Huancayo (HUA: 12.06$^\circ$S, 75.21$^\circ$W geographic; magnetic dip 0.3$^\circ$N) in the Peruvian longitude sector and Alibag (ABG: 18.64$^\circ$N, 72.87$^\circ$E geographic; magnetic dip 10$^\circ$N) in the Indian longitude sector. Six strong geomagnetic storm events during the period 2000-2005, falling in the maximum-to-declining phase of solar cycle 23, are analyzed for this study. These stations are part of the SuperMAG network and data from these stations are openly available. It is noted that although $\Delta$X variations over these two stations are, in general, uncorrelated, significant correlations are observed on certain occasions. Correlation coefficient of at least 75$\%$ for 10 minutes is defined as a requisite criterion to infer the possible connection between the $\Delta$X variations over these two stations. The ionospheric convection maps from the SuperDARN network are also used to understand the Spatio-temporal evolution of the two-cell ionospheric convection patterns over high-latitudes during these periods of observations. This exercise reveals that the $\Delta$X variations over the antipodal locations are significantly correlated when the two-cell convection maps show appropriate rotations and both the stations possibly come under the single cell. Therefore, this investigation brings out a novel method to identify the IMF $B_y$ influence over the low/equatorial latitudes based on the openly available data.

\end{abstract}


\end{frontmatter}


\section{Introduction}

The frozen-in Interplanetary Magnetic Field (IMF) carried by the solar wind, emanating continuously from the Sun, interacts with the terrestrial Magnetosphere-Ionosphere (MI) system forming the basis of the solar-terrestrial interactions. The amount of solar wind energy that would get transferred into the magnetosphere, via the process of magnetic reconnection, is governed by the north-south ($B_z$) component of the IMF \citep{sc:1}. The solar wind-MI coupling becomes strongest when the orientation of the north-south component ($B_z$) of IMF is southward, for a sufficient interval of time, setting conditions favorable for the occurrence of intense/strong (Disturbance storm time: Dst$<=$-100 nT) geomagnetic storms \citep{sc:2,sc:42,sc:3,sc:4,sc:34} thus perturbing the ambient ionosphere over low-to-equatorial latitudes by the process of Prompt Penetration of interplanetary Electric Field (PPEF), causing detrimental effects (see \cite{sc:41,sc:49} and references therein) on the performance of space-based navigation satellite systems like the GPS/GNSS over these locations. 

Although the pronounced and dominant effects of IMF $B_z$ over low-to-equatorial latitudes are known and studied rigorously, there are sparse studies on how the east-west component ($B_y$) of IMF plays a vital role in determining the polarity of PPEF over these locations. Recently, \citep{sc:22} in their work investigated how the changes in IMF $B_y$ would modify this PPEF polarity under steady southward IMF $B_z$. As an extension of the results shown in that paper, the present work makes an effort in investigating whether a strong correlation of the $\Delta$X component of magnetic field (utilizing the SuperMAG data service discussed in the following section) for a sufficient interval of time between two nearly antipodal low-latitude stations: Huancayo (HUA) in the Peruvian longitude sector and Alibag (ABG) in the Indian longitude sector is indicative of the consequence of changing IMF $B_y$ conditions. The analysis has been performed by observing six cases of strong geomagnetic storms that had occurred during 2000 through 2005, falling in the maximum-to-declining phase of the solar cycle 23. The ionospheric convection maps from the SuperDARN (discussed in the following section) are used as evidence of cell rotation of the two-cell plasma convection pattern at the high-latitudes. This study consolidates the physics of the same polarity of penetration electric field at low-latitudes, under southward IMF $B_z$ and varying IMF $B_y$ orientation (eastward and/or westward), thus bringing out the influence of the IMF $B_y$ component at the low-to-equatorial latitudes during geomagnetically disturbed conditions. This study brought that, the role played by IMF $B_y$ in controlling the global ionospheric electrodynamics under geomagnetically disturbed conditions, is a very important field of study, and incorporation of its effects becomes extremely crucial for the development of a reliable space weather forecast system, especially in the low-to-equatorial ionosphere.     

\section{Data}

The SuperMAG service, a global collaboration of national agencies and organizations, operates presently more than 300 ground-based magnetometers. It provides ground magnetic field variations from all the available stations in a common coordinate system, with a common baseline removal approach and with the same time resolution. It utilizes the three-dimensional vector measurements of the magnetic field from these magnetometers. SuperMAG resamples raw data into one-minute temporal resolution and then converts all the units into nT. This data is then rotated into a local magnetic field coordinate system and the baseline is subtracted by an automated technique. See the articles by \cite{sc:25,sc:26} for further details on SuperMAG. 
The Super Dual Auroral Radar Network (SuperDARN) consists of more than 30 low-power high-frequency radars. They observe the Earth's upper atmosphere from the mid-latitudes to the polar regions. These radars are continuously operational to provide information on the near-earth space environment by observing the plasma in the ionosphere. See the review article by \citep{sc:27} for more details on SuperDARN.

For the present study, the $\Delta$X component of the magnetic field measured by the SuperMAG (freely available at http://supermag.jhuapl.edu/) at Huancayo (HUA) of the Peruvian sector and Alibag (ABG) of the Indian sector have been utilized. The magnitude and direction of the IMF $B_y$ for complete southward IMF $B_z$ (Data of the IMFs are available at https://omniweb.gsfc.nasa.gov/form/omni\_min.html) have been investigated to understand the IMF conditions during the analysis period. The ionospheric convection maps, for the evidence of cell rotation during the analysis period, have been used from the SuperDARN network website (available at http://vt.superdarn.org/). It is important to note that HUA (dip: 0.3$^\circ$N) is an equatorial location and ABG (dip: 10$^\circ$N is) is an off-equatorial location. \cite{sc:22} in their work (see Figure 3 of that paper) have shown anti-correlation between the $\Delta$X variations over the respective high- and mid-latitude stations in the Indian and Jicamarca longitude sectors and as one approached low latitudes, this anti-correlation turned into correlation during the interval of the influence of IMF $B_y$ under southward IMF $B_z$ condition. In addition, $\Delta$X variations between ABG and Tirunelveli (TIR; dip: 0.2$^\circ$N), an equatorial location in the Indian longitude sector also showed correlation during this time. Therefore, in this work, we use the readily available ABG $\Delta$X data, which is a low-latitude station, and compare the variations with the corresponding $\Delta$X variations over HUA which is nearly antipodal location of ABG for six events. It is to be noted that the six events under consideration in this work are from the study by \citep{sc:11} where in addition to IMF $B_z$, the importance of the role played by IMF $B_y$ in the process of PPEF in predicting a better forecast lead time of post-sunset equatorial ionospheric irregularities from the sharp northward-southward transition of IMF clock angle, has been clearly shown. Using these six events, further investigations are carried out that explore the efficacy of strong and stable correlations between the magnetic data over the two selected equatorial/low-latitude stations at antipodal locations.       

\section{Results}

This section discusses the results of the six geomagnetic storm events during 2000-2005 where IMF $B_z$ had been completely southward during the observation period, rendering strong geomagnetic storm-time conditions. Firstly, the events with a strong correlation during an eastward orientation of IMF $B_y$ are presented. Next, an event with a complete westward orientation of IMF $B_y$ and finally an event with no significant correlation is shown.

\subsection{Cases of strong correlation with eastward IMF $B_y$}

\subsubsection{Event of April 06, 2000}

\begin{figure}[ht]
\centering\includegraphics[width=1\linewidth]{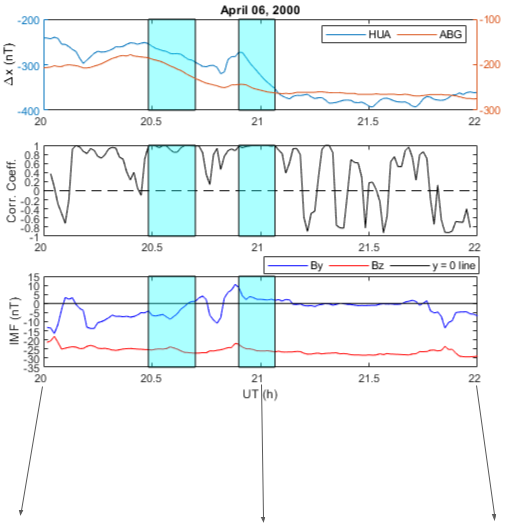}\\
\includegraphics[width=1\linewidth]{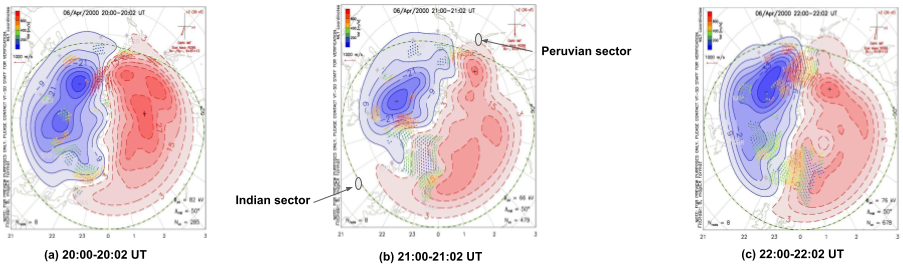}
\caption{Top Panel (Top segment: $\Delta$X component of magnetic field at low-latitude stations HUA and ABG. Middle segment: Correlation plots of $\Delta$X as observed from the two stations. Bottom segment: IMF $B_y$ (in blue) and IMF $B_z$ (in red). Shaded cyan regions mark periods of strong correlation). Bottom Panel: Ionospheric convection maps during (a) 20:00- 20:02 UT, (b) 21:00-21:02 UT and (c) 22:00-22:02 UT of April 06, 2000.}
\label{sc01}
\end{figure}

One of the strongest and well-studied events of solar cycle 23 was the geomagnetic storm of April 06, 2000, caused as a result of a CME that erupted near the western limb of the Sun on April 04, 2000, and hit the Earth's magnetosphere on April 06, 2000, \citep{sc:6,sc:7,sc:8,sc:9,sc:10,sc:11}. Figure \ref{sc01} (top panel: top subplot) shows the variation of the $\Delta$X component of the magnetic field as measured by the magnetometers over two nearly antipodal locations (HUA and ABG) in the low-latitude region. The correlation (top panel: middle subplot) of these magnetic field values over the two locations show a consistent high correlation ($>80\%$ and stable, shaded in the cyan rectangle) between these two stations, that lasted for about 13 minutes from 20:29 UT to 20:42 UT and another for about 10 minutes from 20:54 UT to 21:04 UT. On closely inspecting the IMF variations (top panel: bottom subplot), during the first period of high correlation, the value of IMF $B_y$ had been -6.86 nT (westward) at 20:29 UT, it then turned eastward at 20:38 UT with a value of 0.21 nT and remained eastward at 20:42 UT with a value of 4.10 nT. The corresponding IMF $B_z$ values at these times were -25.10 nT, -26.97 nT and -27.03 nT respectively. During the second period of high correlation, the IMF $B_y$ orientation had been completely eastward with values of 2.13 nT at 20:54 UT and 1.37 nT at 21:04 UT and corresponding IMF $B_z$ values of -24.84 nT and -26.63 nT respectively. Next on observing the bottom panel where the ionospheric convection maps are shown, there had been a significant clockwise rotation of the dawn cell (in red) during (b) 21:00-21:02 UT with respect to (a) 20:00-20:02 UT. An opposite (anti-clockwise) rotation of the dusk cell (in blue) is observed during (c) 22:00-22:02 UT with respect to (b) 21:00-21:02 UT map, as the IMF $B_y$ orientation slowly becomes westward. There had been strong correlations, indicating the possible influence of the same penetration electric field polarity over these two antipodal locations, owing to the clockwise rotation (IMF $B_y$ turning positive or positive during these two intervals) of the cells presumably bringing both the stations within the dawn cell under the influence of IMF $B_y$ in the southward IMF $B_z$ conditions. Furthermore, it is to be noted that the correlation interval from 20:00 UT to 20:20 UT is not considered significant as it does not a show very high and stable correlation unlike the periods from 20:29-20:42 UT and 20:54 to 21:04 UT. Additionally, the westward IMF $B_y$ during this particular period cannot explain the clockwise rotation of the cells bringing both the antipodal stations within the dawn cell.    

\clearpage
\subsubsection{Event of March 20, 2001}

\begin{figure}[ht]
\centering\includegraphics[width=1\linewidth]{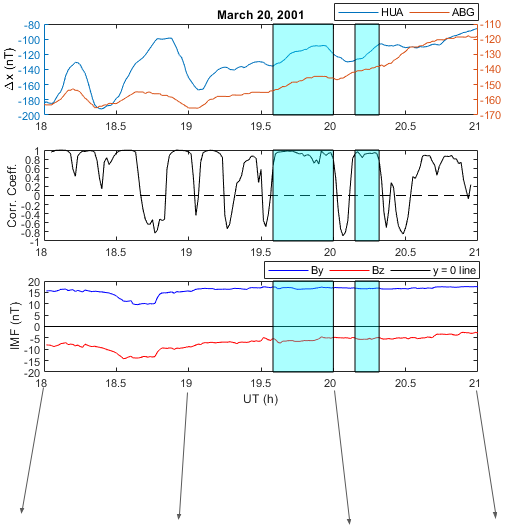}\\
\includegraphics[width=1\linewidth]{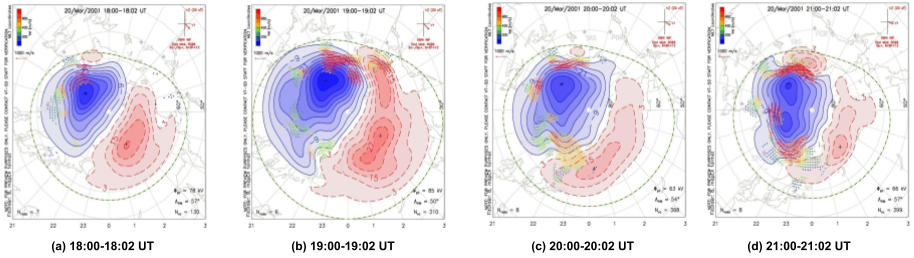}
\caption{Top Panel (Top segment: $\Delta$X component of magnetic field at low-latitude stations HUA and ABG. Middle segment: Correlation plots of $\Delta$X as observed from the two stations. Bottom segment: IMF $B_y$ (in blue) and IMF $B_z$ (in red). Shaded cyan regions mark periods of strong correlation). Bottom Panel: Ionospheric convection maps during (a) 18:00-18:02 UT, (b) 19:00-19:02 UT, (c) 20:00-20:02 UT and (d) 21:00-21:02 UT of March 20, 2001).}
\label{sc02}
\end{figure}

There were observations of two CMEs by the coronagraphs onboard SOHO. The first CME was observed on March 18, 2001, at 02:06 UT and the second CME emerged on March 19, 2001, at 05:26 UT after the collapse of a solar filament (www.spaceweather.com). The former CME was not directed towards the Earth while the latter was Earth-directed which caused a strong geomagnetic storm (G3 class according to NOAA scales) on March 20, 2001. Figure \ref{sc02} shows the magnetic field variations over HUA and ABG and their correlation along with the IMF conditions during the period. Looking into the variation of this correlation, it can be observed that the correlation between the two antipodal stations had been very high (shaded in cyan) for 25 minutes from 19:35 UT to 20:00 UT and for 10 minutes from 20:09 UT to 20:19 UT. During the first period of strong correlation, the values of IMF $B_y$ had been 16.64 nT at 19:35 UT and 17.04 nT at 20:00 UT. The corresponding IMF $B_z$ values were -6.90 nT and -4.95 nT respectively. The second period of strong correlation was present when IMF $B_y$ had a value of 16.63 nT at 20:09 UT with the corresponding IMF $B_z$ value of -5.69 nT and an IMF $B_y$ value of 16.47 nT at 20:19 UT when IMF $B_z$ had a value of -5.55 nT. Thus for this particular case, the IMF $B_y$ values had been consistently around 16-17 nT and were completely eastward. When observing the ionospheric convection maps in the bottom panel, there is a clockwise rotation of the dawn cell during (c) 20:00-20:02 UT with respect to (b) 19:00-19:02 UT.        

\clearpage
\subsubsection{Event of August 18, 2003}

\begin{figure}[ht]
\centering\includegraphics[width=1\linewidth]{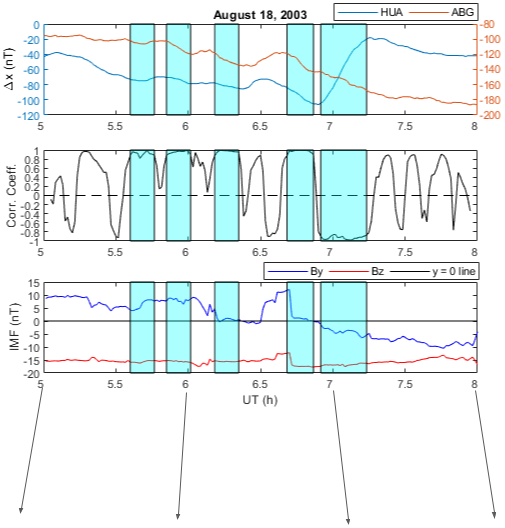}\\
\includegraphics[width=1\linewidth]{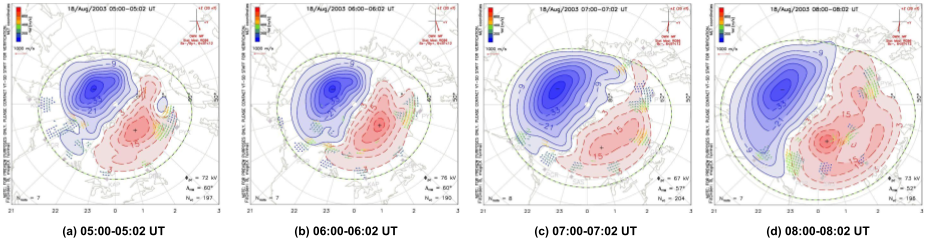}
\caption{Top Panel (Top segment: $\Delta$X component of magnetic field at low-latitude stations HUA and ABG. Middle segment: Correlation plots of $\Delta$X as observed from the two stations. Bottom segment: IMF $B_y$ (in blue) and IMF $B_z$ (in red). Shaded cyan regions mark periods of strong correlation). Bottom Panel: Ionospheric convection maps during (a) 05:00-05:02 UT, (b) 06:00-06:02 UT, (c) 07:00-07:02 UT and (d) 08:00-08:02 UT of August 18, 2003).}
\label{sc03}
\end{figure}

Strong geomagnetic activity, that caused the northern lights (auroras) to be observed as far down as California in the United States, began around 01:00 UT when the IMF $B_z$ tilted south. The Storm Sudden Commencement (SSC) occurred on August 17, 2003, at 14:21 UT while the Dst peaked to a minimum of -168 nT at 16:00 UT on August 18, 2003, \citep{sc:23}. Approaching in a way similar to the previous cases, from the correlation plots of Figure \ref{sc03}, it can be observed that there had been periods (when the orientation of IMF $B_y$ is eastward) of strong correlations and a period (when IMF $B_y$ is westward) of strong anti-correlation between the two antipodal stations. The correlations had been high for 10 minutes from 05:36 UT to 05:46 UT, 05:51 UT to 06:01 UT and 06:11 UT to 06:21 UT and for 11 minutes from 06:41 UT to 06:52 UT. On the other hand, there was a strong anti-correlation lasting for about 19 minutes from 06:55 UT to 07:14 UT. The orientation of IMF $B_y$ (corresponding IMF $B_z$) had been eastward with the values of 4.29 nT (-15.08 nT) at 05:36 UT and 7.47 nT (-15.54 nT) at 05:46 during the first correlation period, the values were 8.75 nT (-15.31 nT) at 05:51 UT and 8.91 nT (-16.84 nT) at 06:01 UT during the second correlation period. The third correlation period showed IMF $B_y$ (corresponding IMF $B_z$) values of 4.20 nT (-15.58 nT) at 06:11 UT and 0.13 nT (15.00 nT) at 06:21 UT while the fourth correlation period showed an eastward orientation of IMF $B_y$ with a value of 1.85 nT (-16.88 nT) at 06:41 UT a value of 0.37 nT (-17.56 nT) at 06:52 UT. The period of strong anti-correlation showed a complete westward orientation of the IMF $B_y$ (corresponding IMF $B_z$) with the values of -2.91 nT (-16.81 nT) at 06:55 UT and -6.70 nT (-16.13 nT) at 07:14 UT. Observations of the convection maps reveal clockwise rotation of the dawn-cell during (b) 06:00-06:02 UT with respect to (a) 05:00-05:02 UT and an anti-clockwise rotation of the dusk-cell during (d) 08:00-08:02 UT with respect to (c) 07:00-07:02 UT.     

\clearpage
\subsubsection{Event of May 30, 2005}

\begin{figure}[ht]
\centering\includegraphics[width=1\linewidth]{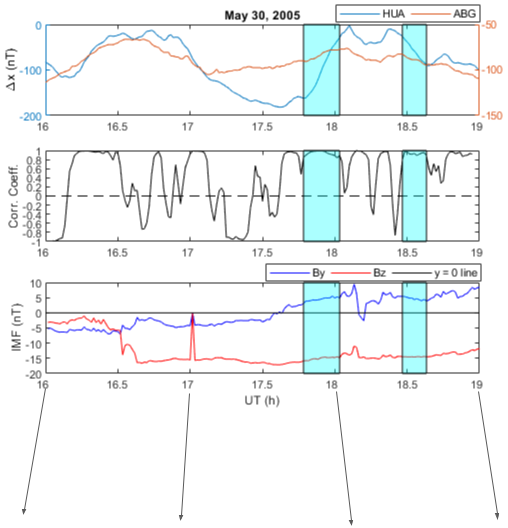}\\
\includegraphics[width=1\linewidth]{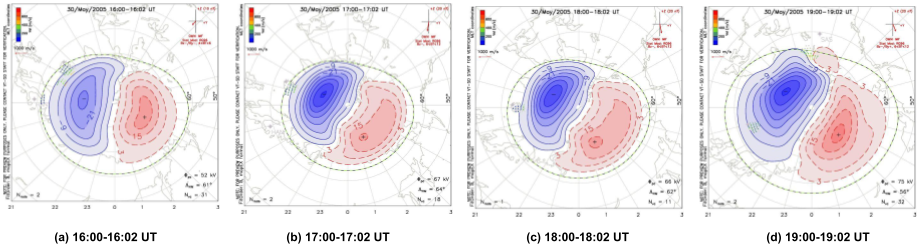}
\caption{Top Panel (Top segment: $\Delta$X component of magnetic field at low-latitude stations HUA and ABG. Middle segment: Correlation plots of $\Delta$X as observed from the two stations. Bottom segment: IMF $B_y$ (in blue) and IMF $B_z$ (in red). Shaded cyan regions mark periods of strong correlation). Bottom Panel: Ionospheric convection maps during (a) 16:00-16:02 UT, (b) 17:00-17:02 UT, (c) 18:00-18:02 UT and (d) 19:00-19:02 UT of May 30, 2005).}
\label{sc04}
\end{figure}

A strong geomagnetic storm had occurred on May 30, 2005, when a CME hit the magnetosphere of the Earth at 09:30 UT on the previous day (see \cite{sc:11} and references therein). Similarly approaching the event as in the previous cases, the two antipodal stations had periods of high correlation about 15 minutes from 17:47 UT to 18:02 UT, and about 10 minutes from 18:48 UT to 18:58 UT, as clearly observed from Figure \ref{sc04} top panel. The first period of strong correlation showed eastward orientation of IMF $B_y$ with values of 3.84 nT at 17:47 UT and 6.97 nT at 18:02 UT with the corresponding IMF $B_z$ values of -15.58 nT and -13.25 nT respectively. The second period of strong correlation showed eastward IMF $B_y$ values of 6.84 nT at 18:48 UT and 8.45 nT at 18:58 UT with the corresponding IMF $B_z$ values of -12.93 nT and -11.79 nT respectively. Observations of the convection maps in the bottom panel show that there had been a clear clockwise rotation of the dawn cell during (c) 18:00-18:02 UT with respect to (a) 16:00-16:02 UT map when the IMF $B_y$ turned eastward. The slight clockwise turning of the dawn cell during (b) 17:00-17:02 UT with respect to the (a) 16:00-16:02 UT map could not be explained with respect to the correlation plots on top of the IMF $B_y$ being westward during that period.       

\clearpage
\subsection{A case of complete westward IMF $B_y$: event of June 18, 2003}

\begin{figure}[ht]
\centering\includegraphics[width=1\linewidth]{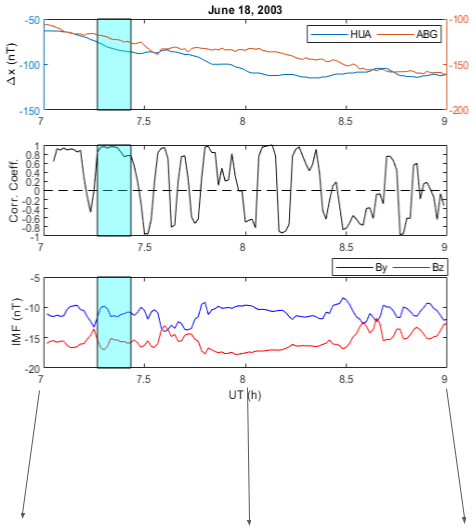}\\
\includegraphics[width=1\linewidth]{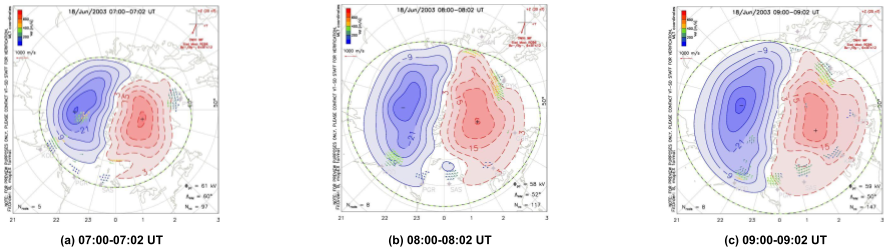}
\caption{Top Panel (Top segment: $\Delta$X component of magnetic field at low-latitude stations HUA and ABG. Middle segment: Correlation plots of $\Delta$X as observed from the two stations. Bottom segment: IMF $B_y$ (in blue) and IMF $B_z$ (in red). Shaded cyan regions mark periods of strong correlation). Bottom Panel: Ionospheric convection maps during (a) 07:00-07:02 UT, (b) 08:00-08:02 UT and (c) 09:00-09:02 UT of June 18, 2003).}
\label{sc05}
\end{figure}

Due to an interplanetary shock wave sweeping past the Earth's magnetic field on June 18, 2003, at 05:00 UT when a lopsided CME hurled into space on June 15, 2003, from an X-class flare explosion near a sunspot (www.spaceweather.com), a strong geomagnetic storm was triggered on that day. Unlike the previous cases, it is to be noted that during the analysis period of this particular event, IMF $B_y$ had been completely westward. Furthermore, the correlation of magnetic data from the two locations' observation in Figure \ref{sc05} shows a particular period of about 10 minutes from 07:15 UT to 07:25 UT where the correlation had been high ($>=75\%$) while no significant and stable correlation and/or anti-correlation was observed at any other periods. The values of IMF $B_y$, during this period of strong correlation, had been -10.06 nT with the corresponding IMF $B_z$ value of -16.65 nT at 07:15 UT and -11.35 nT with the corresponding IMF $B_z$ value of -15.35 nT at 07:25 UT. Looking into the convection maps during the period of observation in the bottom panel, the anti-clockwise rotation of the dawn-cell was observed during (b) 08:00-08:02 UT with respect to (a) 07:00-07:02 UT map. Due to the complete westward orientation of IMF $B_y$, the dusk cell rotated anti-clockwise causing the two stations to come under the same cell when the magnetic data showed a significantly high correlation.          

\clearpage
\subsection{A case of no correlation: event of August 30, 2004}

\begin{figure}[ht]
\centering\includegraphics[width=1\linewidth]{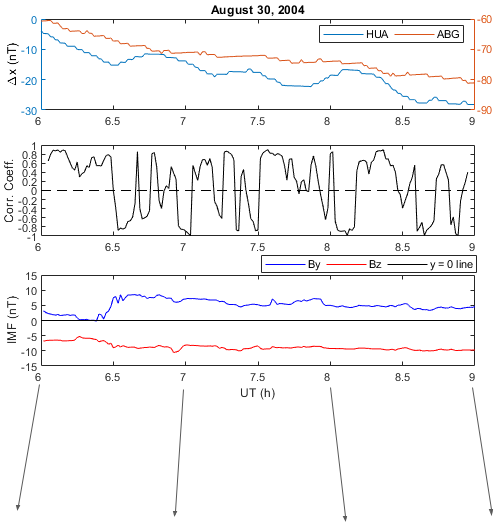}\\
\includegraphics[width=1\linewidth]{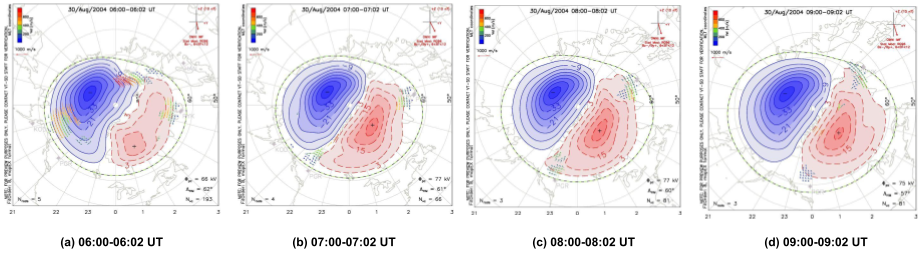}
\caption{Top Panel (Top segment: $\Delta$X component of magnetic field at low-latitude stations HUA and ABG. Middle segment: Correlation plots of $\Delta$X as observed from the two stations. Bottom segment: IMF $B_y$ (in blue) and IMF $B_z$ (in red). Bottom Panel: Ionospheric convection maps during (a) 06:00-06:02 UT, (b) 07:00-07:02 UT, (c) 08:00-08:02 UT and (d) 09:00-09:02 UT of August 30, 2004).}
\label{sc06}
\end{figure}

A geomagnetic storm, accompanied by strong substorm activity, had occurred when there had been an SSC on August 29, 2004, at 10:04 UT. The main phase of the storm was 00:00 UT to 22:00 UT on August 30, 2004, \citep{sc:24}. From Figure \ref{sc06}, it is noted that there had been no substantial-high correlation and/or anti-correlation for a sufficient amount of time. On observing the convection maps in the bottom panel, no significant change in the rotation of the convection cells can be observed. Although the cells stand rotated clockwise throughout during this event, the lack of high and stable correlation between the $\Delta$X variations suggests that the two antipodal stations were not under the same cell for a sufficient interval of time. This is despite the fact that IMF $B_y$ had been completely eastward during the period of observation.

\clearpage
\section{Discussion}

Since the 1970s, several studies \citep{sc:28,sc:20} addressed the utilization of ionospheric measurements for the portrayal of large-scale convection patterns as a function of the IMF. Studies related to the development of statistical models using ground-based high-frequency radar networks and less direct measurements utilizing ground-based magnetometers further consolidated the IMF dependencies of global convection \citep{sc:29,sc:30,sc:31,sc:21,sc:32,sc:33}. The IMF $B_y$ plays a very crucial role in the rotation of the two-cell (DP2) plasma convection pattern (produced by IMF $B_z$) observed in the polar and the high-latitude ionospheric regions \citep{sc:5}. Furthermore, the dayside reconnection geometry gets affected by IMF $B_y$, the magnetospheric convection's dawn-dusk asymmetries are caused as a result of IMF $B_y$ exerting stress on open magnetic field lines and the field-aligned-current distribution gets modulated in the dayside by IMF $B_y$ \citep{sc:12,sc:13,sc:14,sc:15,sc:16,sc:17,sc:18,sc:19}. The amplitude and spatial variation of electric fields in the ionosphere also get modulated due to the influence of IMF $B_y$ \citep{sc:20,sc:21}. Using observations as well as simulations, changes in the high latitude hemispheric asymmetries, thermospheric densities, convection, and wind patterns, under the influence of a changing IMF $B_y$, have also been studied by several researchers \citep{sc:35,sc:36,sc:37}. 

Coming to the works over the low-to-equatorial latitudes, several studies are performed by researchers \citep{sc:44,sc:45,sc:46,sc:47} to establish the crucial role played by IMF $B_z$ in terms of PPEF into the low- and equatorial ionosphere. A study by \citep{sc:48} showed that PPEF has an eastward polarity up to 22:00 LT and becomes westward thereafter during the geomagnetically active periods. However, as the overshielding electric field perturbation has an opposite polarity (westward during daytime and eastward during the nighttime) over the equatorial ionosphere (\cite{sc:41} and references therein), it is common to expect opposite polarities of PPEF during daytime and nighttime at the equatorial locations, consistent with the ionospheric electric field's curl-free condition. This fact is clearly explained in the study by \citep{sc:39,sc:40} where they had found opposite polarities of zonal components of PPEF over the location of Jicamarca in the Peruvian sector and Thumba in the Indian sector, which are nearly antipodal locations. 

As a result of the dominant effects of IMF $B_z$, preventing proper understanding of the effects of IMF $B_y$ during geomagnetically disturbed conditions, it becomes difficult to understand its effects over the equatorial regions. A study by \citep{sc:43} showed the effects of IMF $B_y$ on ring current asymmetry under southward IMF $B_z$. \cite{sc:38} showed modifications in the polarity of the PPEF during geomagnetic storms as a result of variations in the IMF $B_y$ while \citep{sc:22} showed the polarity of PPEF to be the same at nearly two antipodal points (Jicamarca and Thumba) due to the role played by IMF $B_y$ for a particular space weather event of January 7, 2005. \cite{sc:22} observed the evolution (see Figure 3 of that paper) of the $\Delta$X component of magnetic field from showing complete anti-correlation in the high-latitude antipodal locations to becoming strongly correlated in the low-and equatorial antipodal locations, due to the variation in the orientation of IMF $B_y$ that changes the equipotential contour distribution over the equatorial ionosphere. Furthermore, the six events under consideration have been studied by \citep{sc:11} where in addition to IMF $B_z$, the importance of the influence of IMF $B_y$ in the process of PPEF in predicting a better forecast lead time of post-sunset equatorial ionospheric irregularities from the sharp transition of IMF clock angle, has been reported. Taking inspiration from these studies, the present work investigated whether the intervals of the influence of IMF $B_y$ over the equatorial/low-latitude ionosphere be understood from the correlation of the $\Delta$X component of magnetic field variations at nearly antipodal stations (HUA and ABG) in the Peruvian and the Indian longitude sectors. The following subsections briefly discuss the observations of the six events portrayed in the results section.  

\subsection{Cases with strong correlation and eastward IMF $B_y$}

The cases falling under this category were described in Figures \ref{sc01} to \ref{sc04}. From Figure \ref{sc01} a strong rotation of the dawn cell that brought both the antipodal location under the same cell was observed. When IMF $B_y$ started turning eastward and became completely eastward, the convection map of 21:00-21:02 UT showed the aforementioned strong clockwise rotation. As the IMF $B_y$ started to turn westward, the dusk cell rotated anti-clockwise, evident from the 22:00-22:02 UT map, thus bringing back the DP2 configuration in the state comparable with the 20:00-20:02 UT map. As stated earlier, the correlation period from 20:00-20:20 UT was not considered due to the fact that IMF $B_y$ had been westward and this configuration of IMF $B_y$ should have rotated the DP2 anti-clockwise. Further, it did not qualify for the correlation criterion that has been defined. Similar observations were observed for the other three events in Figures \ref{sc02}, \ref{sc03}, and \ref{sc04}. The convection maps in Figure \ref{sc02}, under complete eastward IMF $B_y$ orientation, showed a clockwise rotation of dawn cell evident from the map of 20:00-20:02 UT with respect to the previous map of 19:00-19:02 UT. Here a consistent period (25 minutes) of strong correlation is also observed between the magnetic data of the two stations. another patch of correlation for 10 minutes was also observed suggesting the two antipodal stations were under the same cell for a long duration of time. For the next case, Figure \ref{sc03} revealed periods of correlations during eastward orientation of IMF $B_y$ along with a clockwise rotation of the dawn cell in the map of 06:00-06:02 UT as compared to the previous state. This configuration remained till IMF $B_y$ turned westward when there had been an anti-clockwise rotation of the dusk cell as seen in the map of 08:00-08:02 UT from the previous configuration and a simultaneous strong anti-correlation between the magnetic data was observed during this period. Lastly, on observing Figure \ref{sc04}, it is seen that for eastward orientations of IMF $B_y$, there had been a rotation of the dawn cell as seen from the map of 18:00-18:02 UT with respect to the initial state at 16:00-16:02 UT. Strong correlations of the magnetic data suggest that the two stations were under the same cell during these periods. Further, the slight rotation in the dawn cell as seen in the map of 17:00-17:02 UT with respect to 16:00-16:02 UT, could not be explained as the IMF $B_y$ had been westward during that period, in addition to observation of no significant correlation of the magnetic data between the two locations.       

\subsection{The case with strong correlation under complete westward IMF $B_y$ and the case of no correlation}

In contrast to the previous four cases, Figure \ref{sc05} shows the case when IMF $B_y$ had been completely westward. The period of strong correlation selected could validate the clear anti-clockwise rotation of the dusk cell observed in the map of 08:00-08:02 UT with respect to the previous map, and suggest that the two antipodal locations were under the same cell during this period. The final Figure \ref{sc06} showed no rotations in the dawn or the dusk cell along with no significant correlation period that could qualify for the defined criterion. This case further solidified the fact that the two stations could not be under the same cell and hence no significant correlations were observed during the entire period.  

\section{Summary}

In the present work, a criterion for strong and consistent correlation (having coefficient of at least 75$\%$ for 10 minutes) between the $\Delta$X component of magnetic field (obtained from SuperMAG) at two equatorial/low-latitude stations (HUA and ABG) at nearly antipodal locations in the Peruvian and Indian sectors respectively, has been defined. For the analysis purpose six geomagnetic storm events during the period 2000 through 2005, falling in the maximum-to-declining phase of the solar cycle 23, are analyzed. Under certain occasions, when this criterion is getting satisfied, significant correlations are observed over these two stations, which are generally uncorrelated. The SuperDARN ionospheric convection maps are used to understand the spatial and temporal evolution of high-latitude DP2 convection patterns. Observations of clockwise rotation of the dawn cell during eastward IMF $B_y$ and anti-clockwise rotation of the dusk cell during westward IMF $B_y$, when the criterion of correlation was getting satisfied, are brought forward from the events. The event of August 30, 2004, validates this criterion further where there had been no significant correlation of the magnetic data along with no rotation of the DP2 convection pattern. These observations reveal that the magnetic data over these antipodal locations are significantly correlated when the DP2 pattern has shown appropriate modification in sizes and rotations, thus bringing the two stations under a single cell. This work brings out a novel method, using openly available data, in the identification of the influence of IMF $B_y$ over the low/equatorial locations under southward IMF $B_z$ conditions, thus making the same useful for consideration and development of a reliable space weather forecast system, especially over these geosensitive locations.      

\clearpage
\section*{Acknowledgments} 

The authors acknowledge the use of magnetic data openly available from the SuperMAG network website (http://supermag.jhuapl.edu/). The authors further thank the PIs of the magnetic observatories and the national institutes that support the observatories. The authors also acknowledge the use of SuperDARN data openly available from the Virginia Tech SuperDARN website (http://vt.superdarn.org/).  SuperDARN is a collection of radars funded by national scientific funding agencies of Australia, Canada, China, France, Italy, Japan, Norway, South Africa, the United Kingdom, and the United States of America. Further acknowledgments go to NASA OMNIWEB Data Explorer website (https://omniweb.gsfc. nasa.gov/form/omni\_min.html) for the high resolution (1 minute) IMF $B_y$ and IMF $B_z$ data. This research work is supported by the Department of Space, Government of India.

\clearpage
\bibliographystyle{model2-names.bst}\biboptions{authoryear}

\bibliography{PDF-P1.bib}

\end{document}